# A Selective Metasurface Absorber with An Amorphous Carbon Interlayer for Solar Thermal Applications

Chenglong Wan[1], Yinglung Ho[1], S.Nunez-Sanchez[1], Lifeng Chen[1], M.Lopez-Garcia[1], J.Pugh[1], Bofeng Zhu[1], P. Selvaraj[2], T.Mallick[2], S.Senthilarasu[2] and M.J.Cryan[1]

[1]Department of Electrical and Electronic Engineering, University of Bristol, Bristol, BS8 1UB, UK

[2]Environment and Sustainability Institute, Penryn Campus, University of Exeter, Cornwall, TR10 9FE, UK

E-mail: m.cryan@bristol.ac.uk and cw7679@bristol.ac.uk

## Abstract

*This paper presents fabrication, measurement and modelling results for a metal-dielectric-metal metasurface absorber for solar thermal applications. The structure uses amorphous carbon as an interlayer between thin gold films with the upper film patterned with a 2D periodic array using focused ion beam etching. The patterned has been optimised to give high absorptance from 400-1200nm and low absorptance above this wavelength range to minimise thermal radiation and hence obtain higher temperature performance. Wide angle absorptance results are shown and detailed modelling of a realistic nanostructured upper layer results in excellent agreement between measured and modelled results. The use of gold in this paper is a first step towards a high temperature metasurface where gold can be replaced by other refractory metals such as tungsten or chrome.*

## Introduction

Selective surfaces have reflectance spectra that are tailored to a particular application [1], [2]. For solar thermal applications an ideal surface would absorb all energy across the solar spectrum and simultaneously need to prevent as much thermal radiation as possible in order to maximise the operating temperature. From Kirchoff's law of thermal radiation if a surface has zero absorption there is no emission of radiation. Thus a solar thermal selective surface would have very high absorption across the solar spectrum up to a wavelength at which thermal radiation becomes significant. A number of technologies have been explored for selective absorbers such as multilayer nanorod absorbers [3], metamaterials [4], ceramic-metal composites (cermets) [5] and photonic crystals [6]. Multilayer absorbers usually use alternating layers of different materials, typically dielectric and metal, that produce multiple reflections within the structure which can be used to tailor the absorption spectra. The bottom reflecting metal layer has high reflectance in the infrared (IR) region and top dielectric layer reduces the visible reflectance [7]. Cermets consists of fine metal particles in a dielectric or ceramic matrix, or a porous oxide impregnated with metal [8]. These films are transparent in the thermal infrared region, but are strongly absorbing across the solar spectrum because of interband transitions in the metal and the small particle resonances. When the cermet is deposited on a highly reflective mirror, the tandem forms a selective surface with high solar absorptance and low thermal emittance [9]. Metal-dielectric-metal (MDM) structures are an example of a metamaterial which have a patterned metal layer on a dielectric interlayer on a lower metal [10], [11]. The patterned metal-dielectric-metal structure enables a balance between electric and magnetic resonances to be formed within the structure which results in broadband impedance matching to free space, while the periodicity is small enough such that the structure is strongly reflecting in the IR [10].

We recently showed modelling results for a metasurface absorber based on an MDM structure [11] that has an amorphous carbon interlayer that produces very good selective surface performance with wide angular response. The reasons for using an amorphous carbon layer are twofold. Firstly we are considering applications in solar thermionic energy converters [12], [13] which use hot lithiated diamond surfaces as the source for thermionically emitted electrons. If a diamond layer could be integrated within a selective surface this could lead to a very low cost implementation of a solar thermionic device. Here, the use of amorphous carbon is an initial step in that direction. Secondly, using strongly absorbing materials in a selective surface would appear to be an interesting research direction which has only received limited attention to date [14]. Thus in this paper we fabricate and characterise a low temperature MDM selective surface based on a gold-carbon-gold tri-layer structure and analyse its angular optical response, which is very important in solar energy applications.

Figure 1 shows a schematic of the optimised structure with dimensions obtained from our previous paper [11]. The tri-layer structure is relatively straightforward to obtain using standard deposition techniques. The challenging part of the fabrication process is the nanoscale patterning of the upper layer. In this paper we use Focused Ion Beam (FIB) etching to create a ~20x20µm structure which can be characterised using our in-house Fourier microscope.

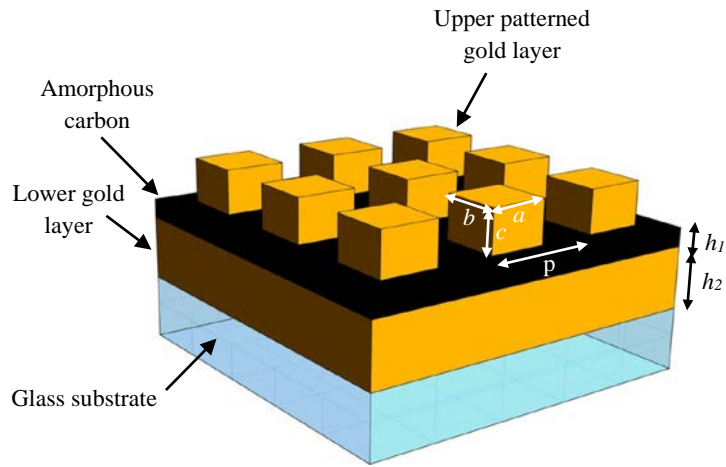

*Figure 1: 3D schematic view for MDM tri-layer with upper gold patterning : a=b=105nm, c=100nm, p=175nm, amorphous carbon, $h_1$ = 35nm, lower Au, $h_2$ = 300nm as shown in* [11]*, on 1mm glass*

The paper is organised as follows: The fabrication steps for the tri-layer are discussed in detail along with a FIB image of a cross-section showing resulting layer thicknesses. Then FIB etching of the upper 2D patterned layer is described. In particular the role of dwell time and beam current are discussed along with resulting FIB images of the structures. Optical characterisation is then carried out using both Fourier microscopy (FM) and Fourier Transform InfraRed (FTIR) spectrophotometry. Reflectance, transmittance and absorptance measurements of amorphous carbon films on glass, amorphous carbon on gold and MDM selective surfaces have been performed and are compared with Finite Difference Time Domain (FDTD) modelling.

# Fabrication

The MDM structure was fabricated on a 24mm x 24mm x 1mm glass substrate. The gold was deposited by sputter coating (Leybold L560) at a fixed deposition rate of 9 Angstroms/sec. An ultra-thin Titanium layer (~10nm) was used as an intermediate layer to improve gold adhesion [15]. The carbon inter-layer was deposited by a Modular Coating System (Quorum Technologies Q150T) directly on top of the gold layer. The coating system used pulse evaporation from a spectographically pure carbon rod with impurities <2ppm. The system settings were : pulse current = 65A, pulse length = 5.00 s and number of pulses = 3. The coating thickness is determined by the amount of carbon rod used which can be carefully controlled and the deposited layer thickness is accurately measured by an in-situ film thickness monitor.

All metasurfaces have been fabricated on the same optimised MDM tri-layer sample. Figure 2 shows a cross-section of this tri-layer structure viewed after FIB cross-sectioning. A thin layer of platinum has been deposited over the upper gold layer to aid the imaging and cross-sectioning process. The dark amorphous carbon layer can be seen clearly between the two gold layers and the titanium adhesion layer can also can be seen between gold and glass. The thicknesses of the layers were measured in 10 sample positions and mean and standard deviation values are shown here : upper gold =115.7±0.3nm, amorphous carbon=36.8±0.4nm, lower gold = 332.7±0.6nm and these are close to our previously optimised strucure as shown in Figure 1.

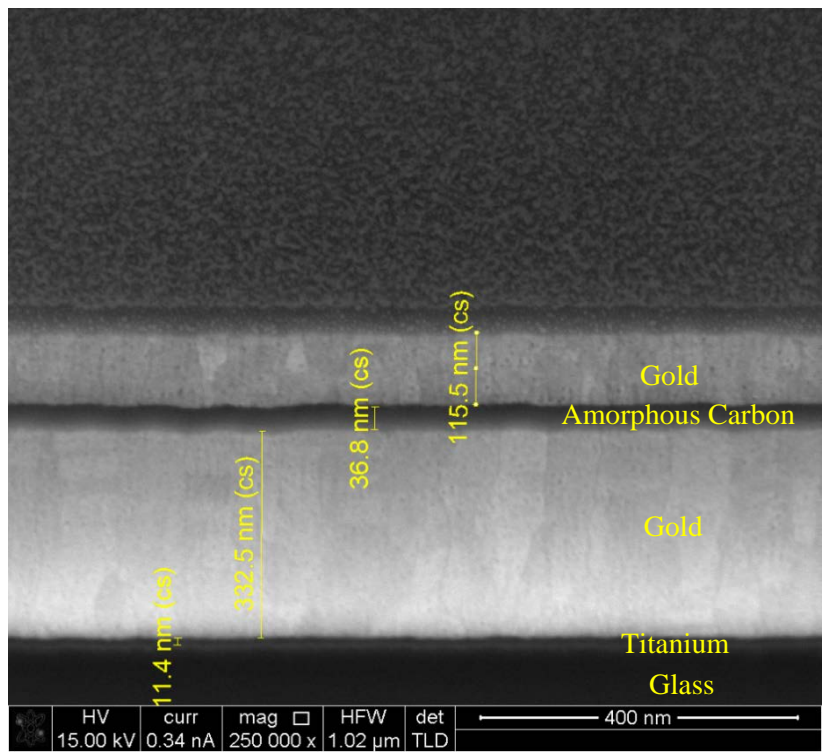

*Figure 2: SEM view of a FIB cross-section of the MDM tri-layer with: upper gold = 115.7±0.3nm, amorphous carbon = 36.8±0.4nm, lower gold = 332.7±0.6nm*

The upper patterned layer was fabricated by FIB etching using a gallium beam with a resolution of approximately 5-10nm. A FEI Strata FIB201 instrument was used with 30keV energy with a chamber pressure of $1.5 \times 10^{-6}$ mbar. The square arrays were created by etching 120 parallel lines of width = 70nm, spaced horizontally by 105nm, this process is then repeated vertically. A single repitition was used with an extended dwell time as discussed below in order to maintain the highest possible resolution. To etch a structure of 20x20μm area takes approximately 2 hours.

The FIB current and dwell time are the main parameters that effect the etching width and depth for a designed nanostructure. It should be noted here that the carbon inter-layer plays an important role in allowing the charged gallium ions to be discharged through the conducting carbon layer. Figure 3 shows SEM images of four different upper patterned layers. Initially we used beam currents of 4pA and 11pA with 480μs dwell time, these are shown in Figures 3(a) (Structure A) and (b) (Structure B), respectively. Figure 3(a) shows that there was incomplete etching through the upper gold layer. However, Figure 3(b) shows that the structure was over-etched if we use 11pA beam current. In Figure 3(c) (Structure C) and (d) (Structure D), we used 11pA beam current again but reduced the dwell time to 450μs and 400μs, respectively. In both cases we obtain uniform and periodic arrays that have almost no interconnecting feature remaining. Figure 3(d) comes closest to the ideal set of parameters described in Figure 1. The average period is 880nm/5 = 176nm which is close to our optimum case (175nm) but each lattice length is approximately 120nmx120nm, ~15nm larger the simulation results (105nm).

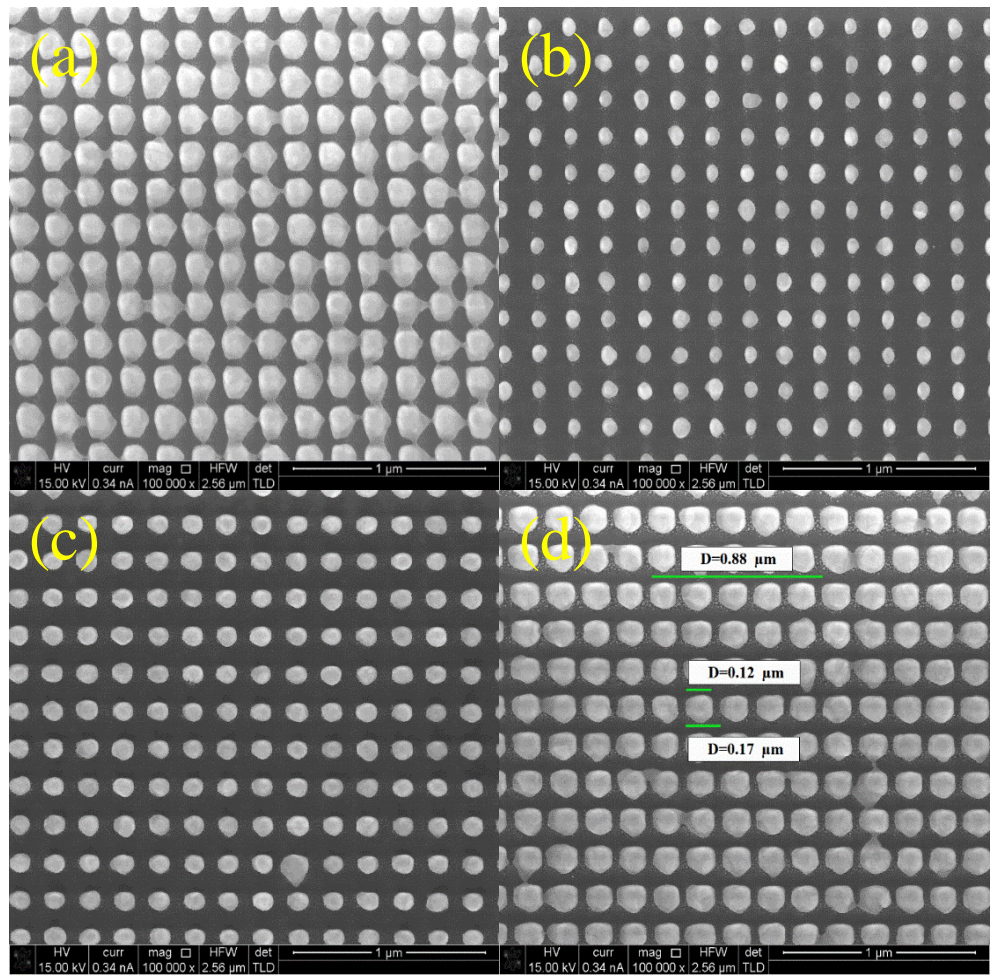

*Figure 3: SEM view of four MDM structures with FIB etched surface patterns (a) 4pA beam current, 480μs dwell time (Structure A); (b): 11pA beam current, 480μs dwell time (Structure B); (c): 11pA beam current, 450μs dwell time (Structure C); (d): 11pA beam current, 400μs dwell time (Structure D)*

## Optical Characterisation

The Reflectance (*R*) and Transmittance (*T*) measurements were obtained using Fourier transform infrared spectroscopy (FTIR) and Fourier microscopy (FM) [16]. The FTIR instrument is a Perkin Elmer 1050 model and the FM system is an in-house setup with an Ocean Optics spectrometer. FTIR is able to measure *R* and *T* at normal incidence for a target area larger than 0.5cm$^2$ and uses a polychromatic light source with a wide wavelength range from 0.25-2.5μm [17]. Because the FTIR system only works with large area samples, we have used the capability of the FM technique to measure the nanopatterned structures which have areas of ~20μm x20μm. The major advantage of FM is that one obtains wide angular reflectance and transmittance data from a single measurement using an objective with a high numerical aperture (Olympus MPLFLN 40X, NA=0.75). A detailed description of experimental set-up is given elsewhere [16]. The Fourier microscope used here uses two different spectrometers for the visible and infrared part of the spectrum and thus results shown here are a composite of these two spectra.

Initially we needed to measure the optical properties of the amorphous carbon and confirm they are close to those used in our modelling work. To do this we deposited ~35nm of carbon on glass and characterised it in the Fourier microscope. The infrared data was smoothed due to low light intensities from the source in the infrared in comparison with the visible range. The real and imaginary parts of the refractive index of the amorphous carbon were shown in our previous paper from the visible up to 10μm [18]. The dash line in Figure 4 shows 2D Finite Difference Time Domain [19] modelled reflectance and transmittance for a 35nm thick amorphous carbon layer on glass. The absorptance (*A*) can be derived from this data from $T + R + A = 1$. The solid line shows experimental results measured by FM from 0.5-0.9μm and 0.9-1.6μm separately. The measured reflectance is very close to the FDTD simulation but transmittance is slightly higher than we expect in the near IR regime.

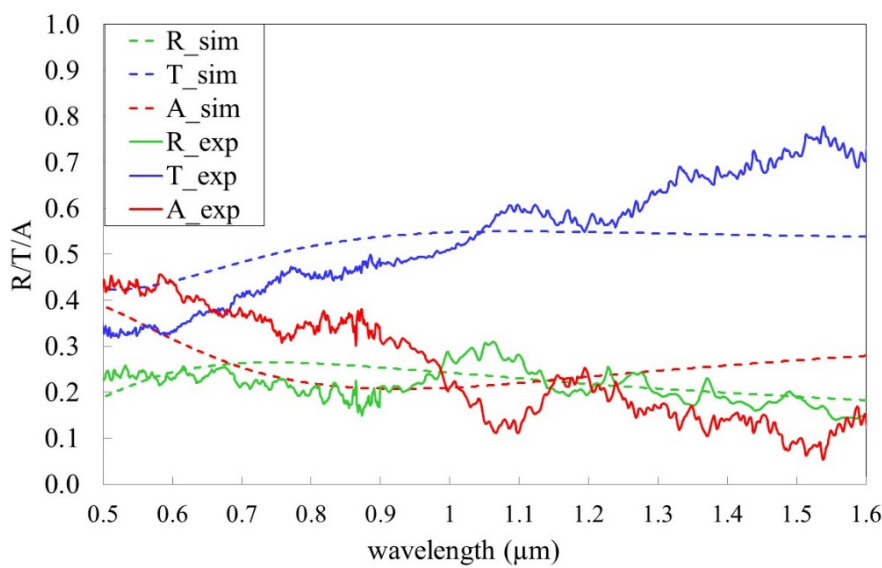

*Figure 4: Experimental Fourier microscope and simulated results for Reflectance (R), Transmittance (T) and Absorbance (A) for 35nm amorphous carbon on a glass substrate at normal incidence*

Figure 5 shows absorptance derived from FTIR measurements for the case of a thick gold layer (~300nm) and a thick carbon layer (~200nm) on gold. The results show that the thick gold layer has strong absorption below 0.5μm and strong reflectance at longer wavelengths. It can also be seen that when a thick carbon layer is deposited on top of the gold, the bi-layer has strong absorptance across the whole spectrum. There is also what appears to be an interference fringe with a period of ~1.5μm which is also observed in the modelled results. The differences between the measured and modelled results for the thick carbon layer on gold could be due to the fact that the carbon layer was produced by multiple depositions. Thus the purpose of our design is to enhance the total absorptance of the bi-layer at shorter wavelengths and extend this absorptance to around 1.2μm whilst maintaining strong reflectance in the infrared. One simple solution is to optimise the thickness of the carbon and this is investigated in Figure 6.

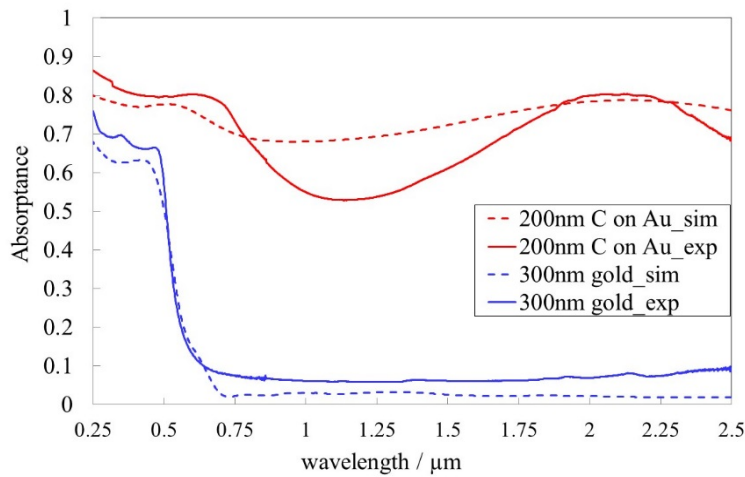

*Figure 5: Experimental FTIR and FDTD simulation results for absorptance of thick carbon layer (200nm) and thick gold layer (300nm) at normal incidence.*

Figure 6 shows results for 25nm, 35nm and 45nm thick carbon layers on 300nm of gold and the absorptance is now much closer to the ideal case discussed earlier. It can be seen that generally good agreement is obtained between the measured and FDTD modelled results. Such thin absorbers placed on metal films have generated some interest recently and can give rise to unexpected effects in the visible wavelength range due to the phase shifts induced at the media interfaces [20]–[22]. At longer wavelengths, where gold becomes more like an ideal metal and the carbon layer thickness becomes a very small fraction of a wavelength, the reflectance of the gold film dominates the behaviour and low absorptance is obtained. It can be seen that good agreement is obtained in the overlapping parts of the spectrum and that absorptance is reasonably independent of incident angle as observed in [20]. These measurements were taken using FTIR, and were repeated using FM over a wavelength range of 0.4-0.85μm and shown in Figure 7. The FM reflectance response shows that the multilayer structure shows an almost angular independence response with a high absorptance, but only below 600 nm.

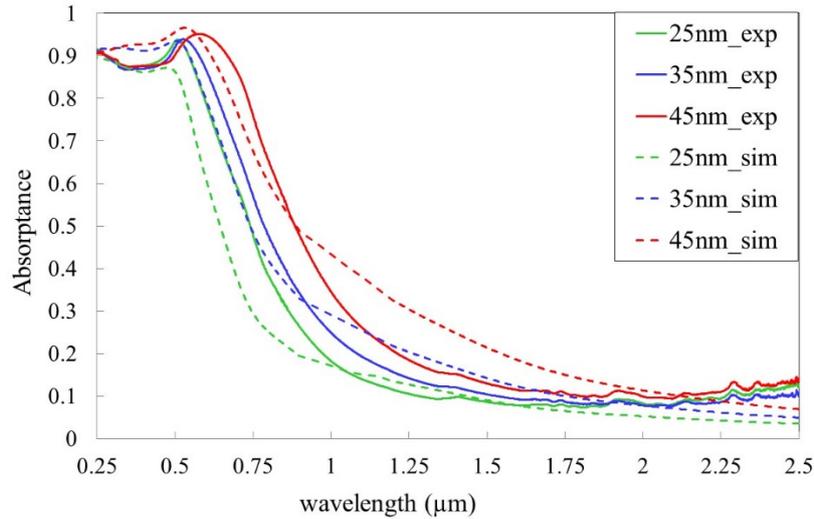

*Figure 6: Experimental FTIR and simulation results of absorptance for different thickness of carbon on ~300nm gold at normal incidence*

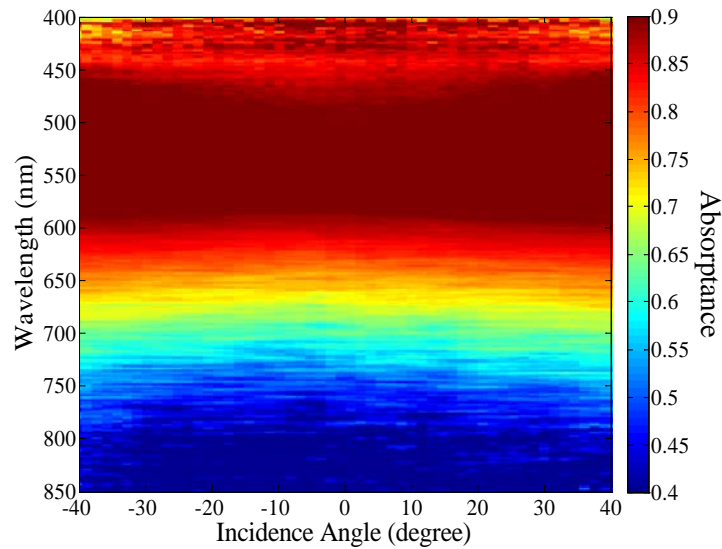

*Figure 7: Experimental FM results of absorptance for 35nm thickness of carbon on thick gold layer vs incidence angle*

The absorptance profile shown in Figure 6 is now much closer to the ideal case, however, if it could be increased between 0.75-1.2μm without increasing the absorptance at longer wavelengths this would further improve the performance. It can be seen in Figure 6 that increasing the carbon thickness improves absorptance around 1μm, but this also tends to increase long wavelength absorptance. Thus a more complex structure is required that allows optimisation of the absorptance below 1.2μm whilst maintaining strong reflectance above this wavelength. The MDM structure allows just such optimisation to be performed. The addition of the upper square array induces a magnetic resonance generated by gap plasmons between the metal layers [23]. This in addition to the electric resonance associated with the upper square array allows careful optimisation of the surface impedance to minimise the reflectance from the structure [24]. The strong absorption in the carbon layer combined with the very high field strengths generated in between the two gold layers results in very strong absorptance at short wavelengths. However at longer wavelengths where these resonances do no occur, the structure is strongly reflective [11].

Figure 8 shows the FM absorptance measurements for both structure A (4pA, 480μs) and structure B (11pA, 480μs) and are compared with the simulated results. As discussed previously in the Fabrication section it is very challenging to fabricate an ideal structure with a uniform pattern with such the small features. Therefore the differences observed at longer wavelengths, where the measured results have lower absorptance than the simulation, could be related to morphology imperfections within the samples. For structure A the large amount of gold remaining between the gaps (see Figure 3(a)) decreases the absorptance of the structure significantly at the longer wavelengths. Once the beam current increases to 11pA in structure B, the etching width is larger than in the designed structure (see Figure 3(b)), which increases the absorptance in the range 0.55μm-0.85μm.

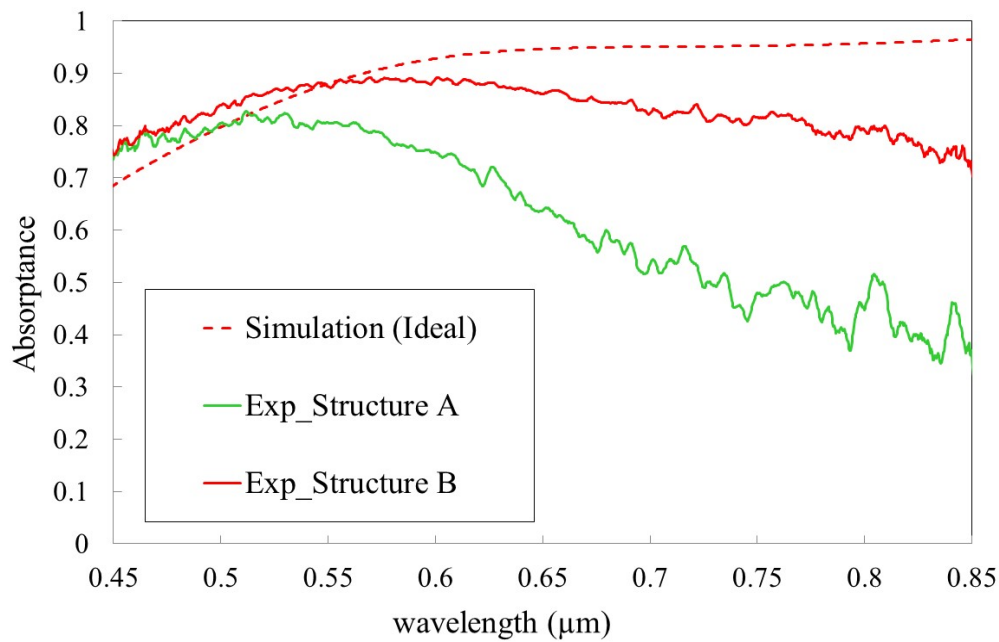

*Figure 8: Experimental FM and simulation results of absorptance for the MDM structure with surface patterns shown in structure A (Beam current = 4pA, Dwell time=480μs) and structure B (Beam current = 11pA, Dwell time=480μs) at normal incidence*

Figure 9 shows the simulation and experimental results for structures C and D. The experimental results were measured by FM from wavelength 0.4-0.85 and 0.85-1.5μm separately. Structure D shows the best selective performance since its aspect ratio is closest to the ideal case. Both structures show strong broadband absorptance up to approximately 90% from 0.4-1.2μm and rapidly reduced to ~30% for structure D and 50% for Structure C at 1.5μm.

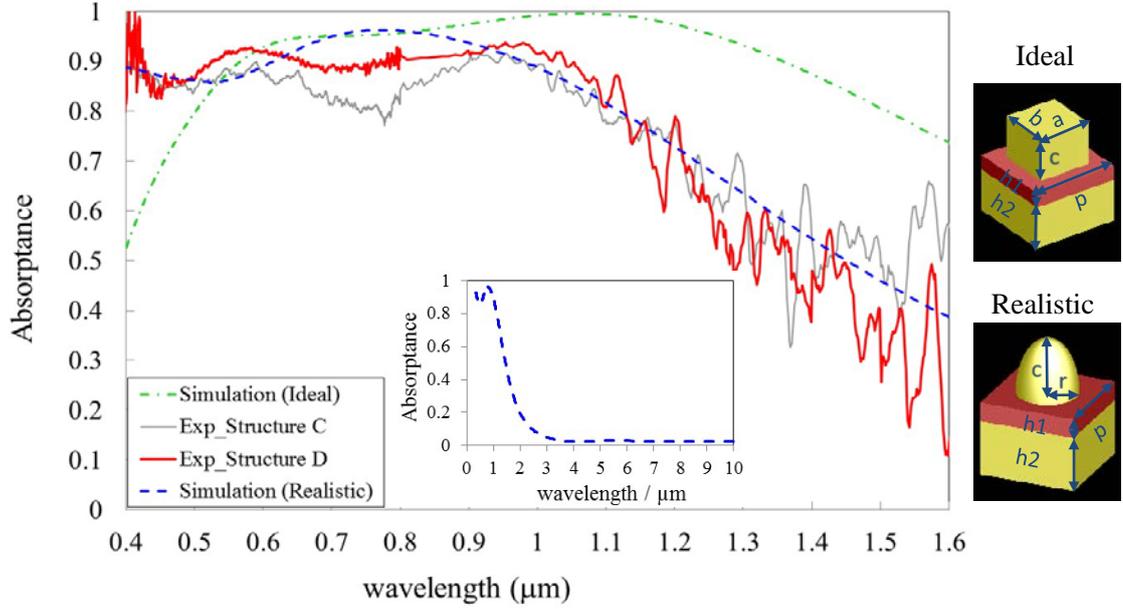

*Figure 9: Simulation (Ideal and Realistic cases as shown on right hand side) and experimental FM results of absorptance for MDM structures C and D at normal incidence, where a=b=105nm, c=100nm, $h_1$=35nm, $h_2$=300nm, r=60nm, p=175nm. Inset shows simulated results for the Realistic case up to 10μm.*

The simulated results for the structure in Figure 1 (Ideal in Figure 9) agree reasonably well with measurements between 0.6μm and 1μm, however, above and below this range the agreement is reduced. Up until this point our FDTD model for the structure has been idealised and we have assumed a simple cubic array for the upper layer. Figure 10 shows both a zoom-out and a zoom-in for Structure D. It can be seen that our fabricated structure is not a perfect cubic array, thus we adapted the structure as shown in Figure 9 to be more realistic. We changed the resonator shape from a cube to half-ellipsoid with height, *c*=100nm and radius, *r*=60nm based on structure D. Figure 9 shows that the simulated response of the realistic structure is in much better agreement at both short and long wavelengths with the experimental response. At shorter wavelengths the increase in absorptance is due to the gradual impedance matching effect of the ellipsoidal geometry. At longer wavelengths the reduction in absorptance is due the change in the aspect ratio from the ideal cubic structure. The main image in Figure 10 also shows that some features still remain connected and this could contribute to the lower measured absorptance at longer wavelengths. Figure 9 also shows simulated results up to 10μm showing that overall the structure has the potential for very good solar selective performance.

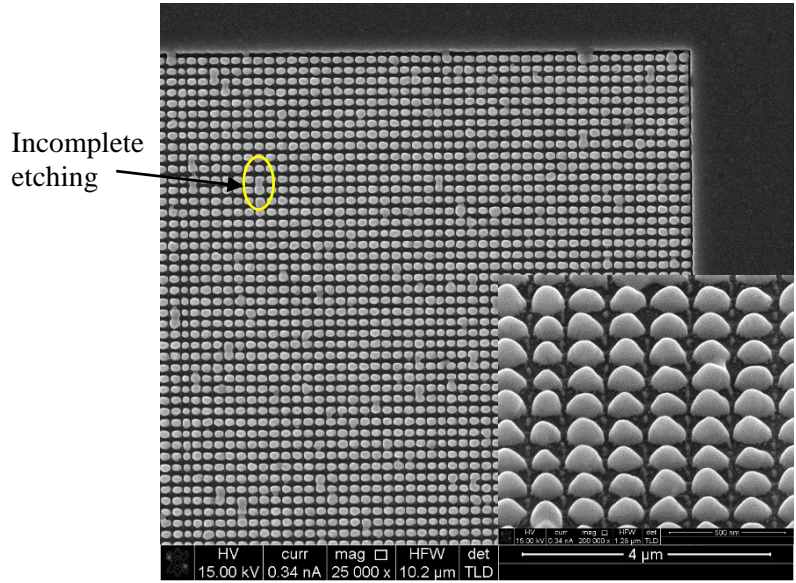

*Figure 10: SEM view of MDM structure D at lower magnification. Inset image at higher magnification showing rounded nature of the upper patterned layer*

In order to verify the performance of our design, a complete absorptance angular response has been measured for the sample with the best response at normal incidence. Figure 11 shows absorptance derived from FM measurements from 0.4µm to 1.6µm in two spectral ranges with incident angles from -40 degrees to +40 degrees. Strong absorptance is obtained from 0.4-1.2µm but strong reflectance from 1.3-1.6µm is shown in part (b). We would like to emphasize the angular independent response of the MDM nanopatterned sample.  The angular independent experimental absorptance of the MDM nanopatterned sample shows there are no diffractive effects due to the presence of a periodic pattern on the top layer. This suggests that absorption is the dominant mechanism in the optical response.  If we compare the optical results in the visible from the multilayer structure (thick carbon layer on top of thick gold layer – Figure 7) and the MDM optimized nanopatterned sample (Figure 11.a), the MDM nanopatterned sample shows a large absorptance in a broader wavelength range than the multilayer sample while maintaining a good angular performance. Work is on-going to measure this structure at longer wavelengths. The main future direction is to move to large area nanopatterning for example using nanoimprint lithography.

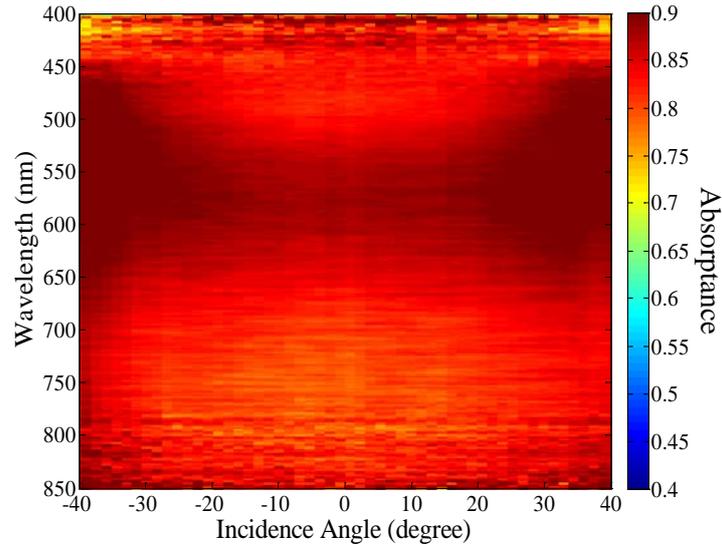

(a)

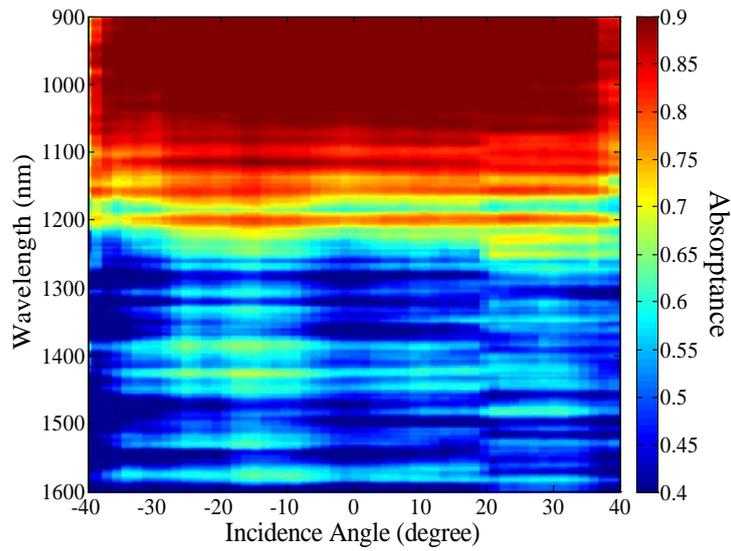

(b)

*Figure 11: Experimental results of absorptance for tri-layer MDM structure with surface patterns show in structure D using 11pA beam current and 400μs dwell time in angular incident angle by the Fourier Microscope from wavelength (a) 400-850nm and (b) 900-1600nm.*

## Conclusions

This paper has presented experimental and FDTD modelling results of ultra-thin amorphous carbon films on glass, on gold and as the interlayer of a metal-dielectric-metal metasurface. The use of a carbon based interlayer not only allows careful optimisation of the absorptance profile, it also allows us to explore the possibility of using these structures in diamond based solar thermionic applications. FIB fabrication was used to create a nanopatterned upper layer to form a MDM metasurface. Careful choice of dwell time and beam current enabled high quality periodic nanostructures to be formed. Both FTIR and FM measurements of the structures have been shown and good agreement with FDTD modelling is obtained. In the case of the MDM structure promising solar selective surface performance is obtained and once a realistic structure was included in the model, excellent agreement is obtained with modelled results. This paper presents a low temperature prototype, in [11] we showed that by replacing gold with a refractory metal such as tungsten, high temperature operation can be achieved in vacuum or in air when a passivating layer such as $SiO_2$ is used.

## Acknowledgements

The authors would like to acknowledge useful discussions with Dr Andy Murray, Dr Peter Heard, Mr Jonathan A Jones and Mr Qiaoyi Wang
The Authors would like to acknowledge the Engineering and Physical Sciences Research Council Grant EP/K030302/1 for funding.
Data shown in this paper is accessible via the University of Bristol data repository: doi:xxx